\documentclass{aa} 

\usepackage{graphicx}
\usepackage{txfonts}
\begin{document}

   \title{Persistent magnetic vortex flow at a supergranular vertex}


   \author{Iker~S.~Requerey\inst{1,2},
          Basilio~Ruiz~Cobo\inst{1,3},
          Milan~Go{\v s}i{\'c}\inst{4,5},
          Luis~R.~Bellot~Rubio\inst{6}
          }

   \institute{Instituto de Astrof\'{i}sica de Canarias, V\'{i}a L\'{a}ctea s/n, E-38205 La Laguna, Tenerife, Spain
                 \and 
                 Max-Planck-Institut f\"{u}r Sonnensystemforschung, Justus-von-Liebig-Weg 3, 37077 G\"{o}ttingen, Germany\\
             \email{requerey@mps.mpg.de}
             \and
                 Departamento de Astrof\'{i}sica, Universidad de La Laguna, E-38206 La Laguna, Tenerife, Spain
                 \and 
             Lockheed Martin Solar and Astrophysics Laboratory, Palo Alto, CA 94304, USA
             \and
             Bay Area Environmental Research Institute, Petaluma, CA 94952, USA
             \and
             Instituto de Astrof\'{i}sica de Andaluc\'{i}a (CSIC), Apdo. de Correos 3004, E-18080 Granada, Spain     
             }

        \titlerunning{Magnetic vortex flow at a supergranular vertex}
        \authorrunning{Requerey et al.}


 
  \abstract
   {Photospheric vortex flows are thought to play a key role in the evolution of magnetic fields. Recent studies show that these swirling motions are ubiquitous in the solar surface convection and occur in a wide range of temporal and spatial scales. Their interplay with magnetic fields is poorly characterized, however.}
   {We study the relation between a persistent photospheric vortex flow and the evolution of a network magnetic element at a supergranular vertex.}
   {We used long-duration sequences of continuum intensity images acquired with \textit{Hinode} and the local correlation-tracking method to derive the horizontal photospheric flows. Supergranular cells are detected as large-scale divergence structures in the flow maps. At their vertices, and cospatial with network magnetic elements, the velocity flows converge on a central point.}
   {One of these converging flows is observed as a vortex during the whole 24\,h time series. It consists of three consecutive vortices that appear nearly at the same location. At their core, a network magnetic element is also detected. Its evolution is strongly correlated to that of the vortices. The magnetic feature is concentrated and evacuated when it is caught by the vortices and is weakened and fragmented after the whirls disappear.}
   {This evolutionary behavior supports the picture presented
previously, where a small flux tube becomes stable when it is surrounded by a vortex flow.}


   \maketitle
%

\section{Introduction}

\begin{figure*}
\includegraphics[width=\textwidth]{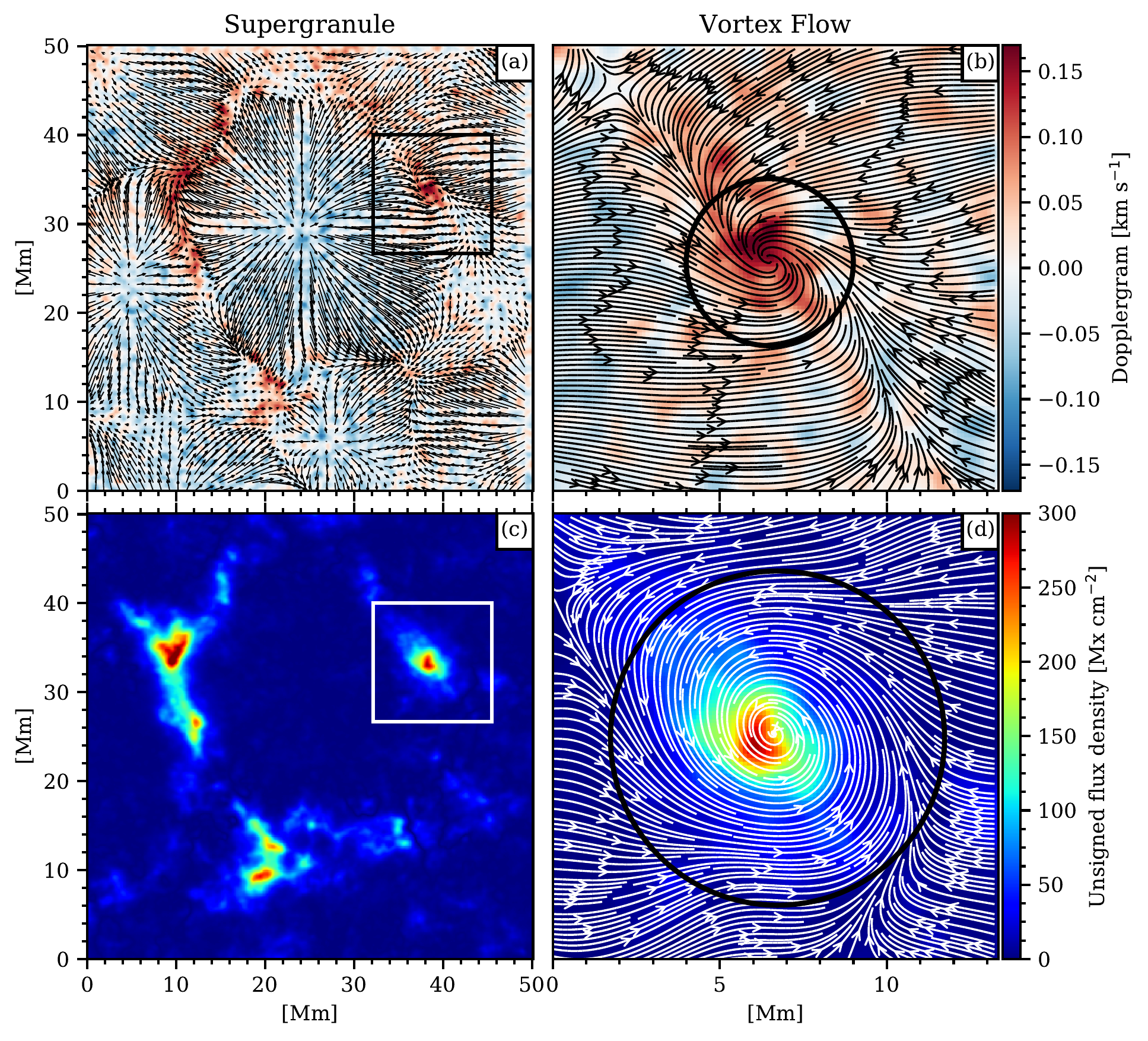}
\caption{Observation of a supergranular cell and a vortex flow in one of its vertices. (a) Quiet-Sun region observed by the \textit{Hinode}/NFI at disk center. The background image displays the mean Dopplergram obtained by averaging over the total duration of the data set, where red and blue regions correspond to downflow and upflow areas, respectively. The black superimposed arrows show the time-averaged horizontal velocity field retrieved with the LCT at the intensity maps. The black solid rectangle, with a FOV of 13.3\,$\times$\,13.3\,Mm$^{2}$, marks the location of the vortex flow. (b) Close-up of the vortex flow. In this case, the black lines display the streamlines of the flows, and the arrows show their directions. Both Dopplergrams are saturated at $\pm 0.17$\,km\,s$^{-1}$. (c) Mean unsigned flux density over the entire FOV. (d) Close-up of the white rectangle in panel
(c). White streamlines display the mean horizontal flow of the magnetogram, i.e., the horizontal proper motions of the NE element that is anchored in the vortex flow.}
\label{fig:fov}
\end{figure*}

In the solar surface, quiet-Sun magnetic fields evolve, driven by convective motions. The magnetic network (NE) outlines the boundaries of supergranular cells \citep{Leighton62,Simon64,Sheeley66}. The cell interiors, the so-called internetwork (IN), are permeated by  horizontal \citep{OrozcoSuarez07,Lites08,Danilovic10a,BellotRubio12} and mixed-polarity \citep{Livingston75,Martin84} magnetic patches located at mesogranular boundaries \citep{DominguezCerdena03,Ishikawa10,YellesChaouche11,Requerey17}. IN elements are advected by supergranular flows from the inner parts of cells toward the edges \citep{OrozcoSuarez12}, thus supplying the total NE flux in only 10\,h \citep{Gosic14}. NE elements consist of vertical kG flux tubes that expand with height \citep[e.g.,][]{Stenflo73,Lagg10,MartinezGonzalez12}. They are very persistent and live for dozens of hours until they are diffused either by flux dispersal or cancellation with nearby elements \citep{Liu94,Hagenaar03,Lamb13}. Magnetic structures are intrinsically unstable against the interchange instability \citep{Parker75,Piddington75},
however, but their lifetimes exceed that of the instability growth time by far.
A stabilizing agent should then exist in order to keep the magnetic elements together for periods of time this long. The likely candidate is a vortex flow around the magnetic feature that is able to stabilize the tube by the centrifugal force of the whirl \citep{Schuessler84}.

Vortex flows are ubiquitous in the solar surface convection. Small-scale vortices with diameters of $\sim$\,1\,Mm were first detected as swirling motions of bright points \citep{Bonet08} and later through local correlation tracking \citep{Bonet10,VargasDominguez11,Requerey17}. Their lifetime varies from 5 to 20\,min \citep{Bonet08,Bonet10}, and they appear located at the vertices of mesogranular cells \citep{Requerey17}. These locations are where the strongest IN magnetic elements are observed to concentrate \citep{Requerey17}. In particular, weak magnetic patches are confined by these converging flows and are later intensified up to kG values as the magnetic element is evacuated \citep{Requerey14}. Larger vortex flows with sizes in the range 5--15\,Mm and lifetimes of at least 2\,h have also been observed at supergranular junctions \citep{Brandt88,Attie09}. \citet{Attie09} found an event where an NE element appears to
be coaligned with a vortex. However, detailed quantitative studies are still needed to confirm this association, and  virtually nothing is known about the influence of vortices on the evolution of NE magnetic fields.

We here study the evolution of a magnetic vortex flow using the Narrowband Filter Imager \citep[NFI;][]{Tsuneta08} on board the \textit{Hinode} satellite \citep{Kosugi07}. The NFI provides very long and stable time series of both continuum intensity images and magnetograms, which are ideal to investigate the evolution of NE elements and their interaction with vortex flows at supergranular vertices.


\section{Observations}

\begin{figure*}
\includegraphics[width=\textwidth]{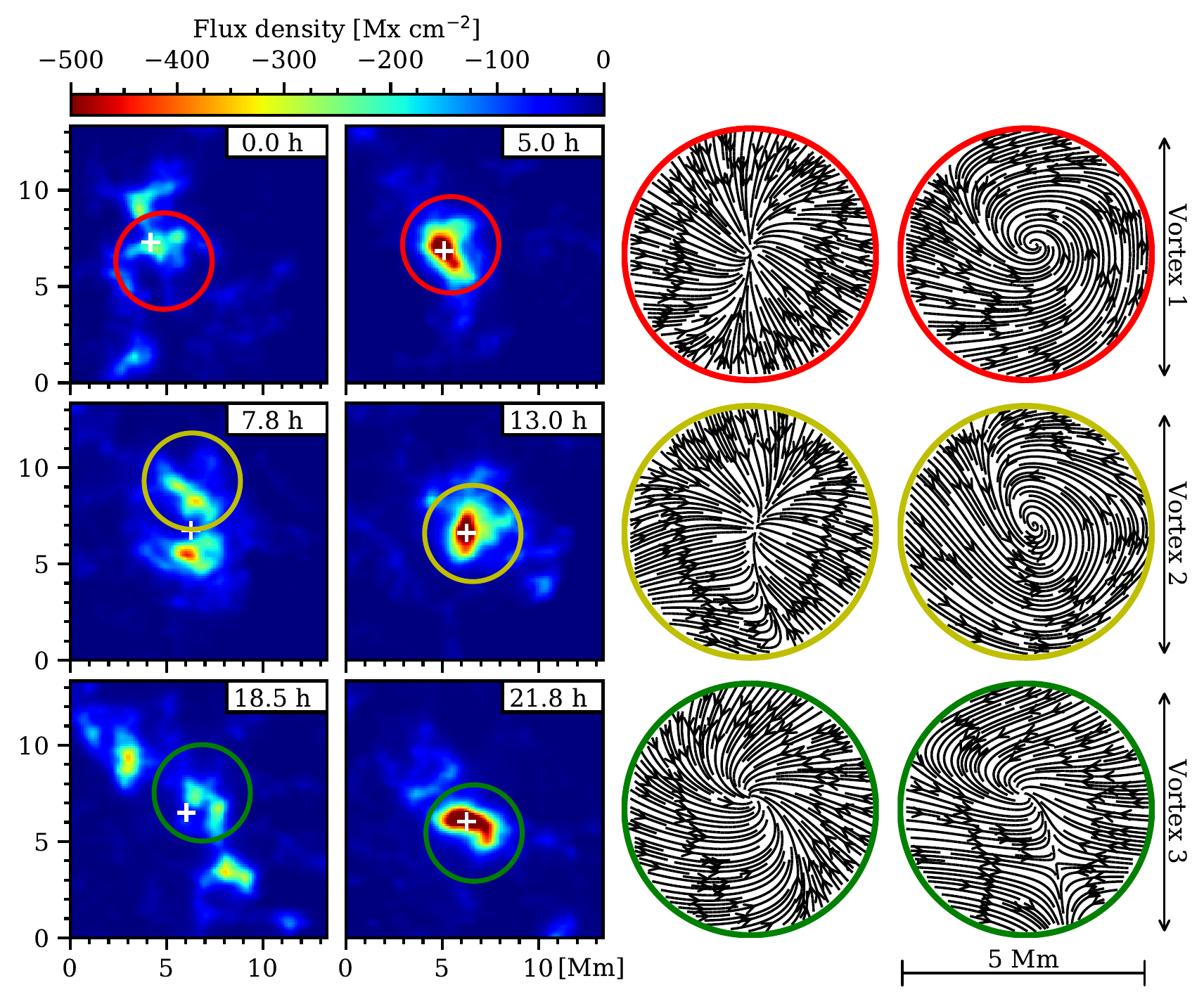}
\caption{Temporal sequence of the magnetic vortex flow. The two left columns display flux density maps. The elapsed time is given in the upper right corner of each panel, and the white crosses show the centroid of the NE structure. The circles with 5\,Mm diameter represent the vortex positions. Their corresponding flow maps are displayed enlarged in the two right columns. Red, yellow, and green circles represent the three different vortex flows that are observed during the time sequence. The temporal evolution is presented in a movie that overlays the magnetic flux density with the streamlines in the whole FOV.}
\label{fig:vortex}
\end{figure*}

\begin{figure}
\includegraphics[width=0.5\textwidth]{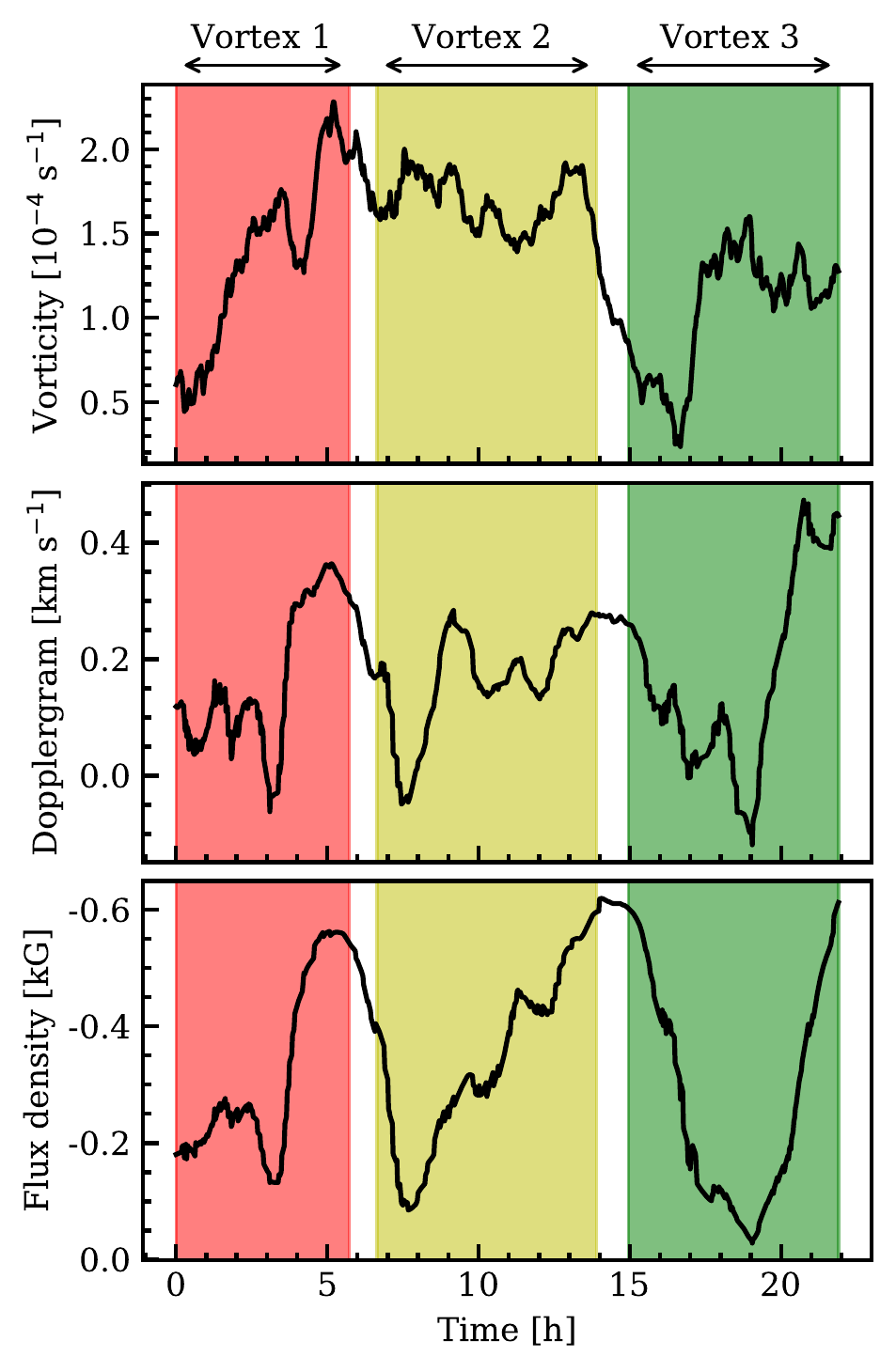}
\caption{Evolution of relevant quantities with full temporal resolution. From top to bottom, the panels show the vorticity, the LOS velocity, and the flux density. The plots display values of the corresponding quantities averaged over a circle with 0.5\,Mm radius centered around the centroid of the NE element.  Red, yellow, and green areas show the time intervals in which the three different vortex flows are observed.}
\label{fig:param}
\end{figure}

The observations used in this paper were obtained with the \textit{Hinode}/NFI on 2010 November 2--3 as part of the Hinode Operation Plan 151. This data set consists of shutterless Stokes I and V filtergrams monitoring the quiet-Sun disk center. The filtergrams were taken in the two wings of the Na\,\textsc{i}\,D1 589.6\,nm line at $\pm 16$\,pm from the line center. From the observed Stokes I and V images, we constructed intensity maps, longitudinal magnetograms, and Dopplergrams, as described by \citet{Gosic14}. The cadence of the observations is only 90\,s, which together with the spatial resolution of about 0\farcs 32 makes this data set perfect for studying the evolution of NE fields. The total duration of the measurements is 38\,h, with two small interruptions of 1\,h and 31\,minutes caused by transmission problems. In this work, we focus on the longest uninterrupted sequence of 24\,h with a total field of view (FOV) of 93\arcsec\,$\times$\,82\arcsec. We reduced the FOV to about 69\arcsec\,$\times$\,69\arcsec (50\,$\times$\,50\,Mm$^{2}$) to focus on a single supergranular cell (see Figure \ref{fig:fov}a). 

The effects of 5-minute oscillations were removed from the intensity maps, magnetograms, and Dopplergrams by applying a subsonic filter \citep{Title89,Straus92} with 5\,km\,s$^{-1}$ as the cutoff. In addition, a 3\,$\times$\,3 Gaussian-type kernel was applied to the magnetograms. This slightly degrades the spatial resolution, but significantly improves the noise level to 4\,Mx\,cm$^{-2}$.


\section{Results}

In order to derive the horizontal flows, we used the local correlation tracking (LCT) technique \citep{November88} as implemented by \citet{Molowny94}. LCT is a powerful cross-correlation technique for measuring proper motions. This technique correlates small local windows in several consecutive images to determine the best-match displacement. The tracking window is defined by a Gaussian function whose full-width at half-maximum (FWHM) is roughly the size of the features to be tracked. In addition, the spatially localized cross correlation is averaged in time to smooth the transition between consecutive images. We applied LCT to the continuum intensity images, Dopplergrams, and magnetograms using a tracking window with an FWHM\,=2.9\,Mm that was averaged over the whole time sequence of 24\,h. 

The mean horizontal velocity fields obtained with the continuum intensity images and the Dopplergrams are almost identical. In Figure \ref{fig:fov}(a) the black arrows display the field obtained with the intensity maps. The background image displays the time-averaged Dopplergram. In the center of the map, a supergranular cell is clearly detected as a large divergence structure. The horizontal velocities are directed from the center of the cell toward its boundary. In the cell interior, the mean Dopplergram displays an upflow, while the boundary is mainly dominated by downflows. A particularly prominent localized downflow in a supergranular vertex is marked by the black solid rectangle. When we zoom in on this area (see Figure \ref{fig:fov}b), the streamlines of the flow show that the downdraft is located at the center of a vortex flow. The vortex has a diameter of $\sim$\,5\,Mm, as marked by the black circle. Photospheric vortex flows like this
at supergranular junctions have previously been detected by \citet{Brandt88} and \citet{Attie09}, but this is the first time that they are related with a downflow, which indicates that they are more likely driven by the “bath tub” effect \citep[cf. subsection 3.3 in][]{Nordlund85}. 

In Figure \ref{fig:fov}(c) we display the mean unsigned flux density over the FOV of \textit{Hinode}/NFI. Owing to the intermittent character of IN elements, only the NE elements are observed after the time-averaging. As expected, the NE flux concentrations are located at the boundary of the supergranular cell, but in particular, they appear to be well localized at sites of strong downflows. One of these NE magnetic elements is located exactly at the center of the vortex flow (see squares in Figure \ref{fig:fov}a and c). Figure \ref{fig:fov}(d) shows an enlarged view of the NE element, with its horizontal proper motions overplotted (white streamlines) as derived from LCT. A counter-clockwise vortex flow is also detected when LCT is applied to the magnetogram. This indicates that the NE element corotates with the vortex. The vortex is larger in the magnetogram than in the intensity map, with a diameter of about 10\,Mm in the former, as indicated by the black circle in Figure \ref{fig:fov} (d). The reason is
that the magnetic concentrations in the magnetograms are larger than their bright point counterparts in the intensity maps.

Vortices have a central vorticity of $1.4 \times 10^{-4}$\,s$^{-1}$ and $1.5 \times 10^{-4}$\,s$^{-1}$ in the intensity map and magnetogram, respectively. This corresponds to a rotation period of about 12\,h, which is on the same order of magnitude as the large vortex found by \citet{Brandt88}, but ten times smaller than those found in IN areas at the junctions of mesogranular cells \citep{Bonet10,VargasDominguez11,Requerey17}. 

\subsection{Evolution of a magnetic vortex flow}

The long duration of our time sequence allows us to study the evolution of the magnetic vortex. To do this, we applied LCT to the intensity maps with a time average of 2\,h. To obtain the evolution of the horizontal velocities, we moved this time window with a step length of one frame throughout the 24\,h data set. We applied the same temporal averaging to the Dopplergrams and magnetograms. The two left columns in Figure \ref{fig:vortex} display the time sequence of the magnetogram. In each panel the circles mark the location of the vortex flow at that moment. Close-ups of their corresponding horizontal flows are shown in the two rightmost columns. The whole evolution is presented in a movie included in the electronic edition of the journal.

In Figure \ref{fig:param} we quantitatively analyze the evolution of the magnetic vortex shown in Figure \ref{fig:vortex}. To do
this, we tracked the NE structure in the magnetograms by computing a region whose summed magnetic flux was constantly equal to $-1.2\times 10^{19}$\,Mx throughout the period of observation. These regions were constructed by starting from the most intense pixels in the magnetogram and then gradually expanding the area by adding the next most intense pixels. After we covered the constant magnetic flux area, we also derived its centroid. These centroids are marked by a white cross in the magnetograms of Figure \ref{fig:vortex}. From top to bottom, the panels of Figure \ref{fig:param} show the evolution of the vorticity, the line of sight (LOS) velocity, and the flux density at the centroid of the NE element. To increase the signal-to-noise ratio of the physical parameters, we represent averages over a 0.5\,Mm radius circle around the centroid.

During the time sequence, three recurrent vortices are observed. Their centers were detected using Lagrange tracers as in \citet{Requerey17}.
In each horizontal velocity map,  we tracked the evolution of individual corks. Initially, a cork is located at each pixel of the image, and then they are advected, driven by the horizontal velocity field. If the corks are tracked long enough, they finally reach the centers of vortex flows where the velocity vectors converge on a central point.  We obtained the central position of the vortex in each frame in this way, and we considered that a vortex survived as long as it shows a convergence center. Each vortex flow is represented by a different color, namely, red, yellow, and green for the first, second, and third vortices, respectively. In each row of Figure \ref{fig:vortex}, we show the NE element at the initial and final stages of each vortex. The time span of each vortex is also shown with the corresponding colors in Figure \ref{fig:param}. The vortex flows last about 7\,h. At the initial phase of each vortex, the NE structure is weak and dispersed, but it becomes concentrated and intensified toward the end of the vortex life. This results in an oscillatory evolution of the NE magnetic field, which shows three recurrent intensification processes (see bottom panel in Figure \ref{fig:param}) associated with the three consecutive vortices. This oscillatory behaviour of the magnetic field was first detected in IN elements \citep{MartinezGonzalez11} and was latter shown to be produced by the buffeting of surrounding granules \citep{Requerey15} and in particular by vortex flows \citep{Requerey17} at the junctions of mesogranular cells.  

In the animation provided as online material, we can observe that the magnetic feature is concentrated when it is caught by the vortices and is weakened and fragmented as soon as the whirls disappear. The evolution of the NE element is correlated not only with that of the vortices, but also with the LOS velocity (see Figure \ref{fig:param}). The downflow increases when the magnetic structure intensifies and decreases just at the moment when the magnetic fields weaken. 

\section{Discussion and conclusions}

The long duration and high spatial resolution of the \textit{Hinode}/NFI Stokes I and V filtergrams have allowed us to detect the most persistent photospheric vortex flow ever observed on the Sun. The mean vortex flow appears to be located at a supergranular vertex that is cospatial with a localized downflow, and it is detected even after averaging over the whole 24\,h time sequence. It has a diameter of 5\,Mm and a rotation period of about 12\,h. These values agree with the large-scale vortices previously reported by \citet{Brandt88} and \citet{Attie09}. In addition, we here also detected an NE structure that corotated with the vortex. 

When we studied the evolution of the vortex using an LCT time average of 2\,h, we realized that the mean vortex flow indeed consists of three consecutive vortices with a lifespan of about 7\,h each. In agreement with the theoretical work of \citet{Schuessler84}, vortices seem to play an important role in stabilizing the NE element. The magnetic structure is concentrated and intensifies when it is caught by a vortex and weakens and disperses as soon as the swirls disappear. The downflow in the core of the magnetic feature also varies in phase with the flux density, suggesting that the tube is evacuated and undergoes convective collapse at each intensification phase \citep[see, e.g.,][]{Parker78,Nagata08,Danilovic10b,Requerey14}. At this point, an interesting question arises: is the downflow associated with the convective collapse of the NE element that
triggers the formation of the vortex through the bath-tub effect \citep{Nordlund85}, or does conversely the vortex perturb the magnetic flux feature and causes it to collapse? Figure \ref{fig:param} shows a delay between the velocity and flux density and  the vorticity. This is better seen in the two local minima at the beginning of the second and third vortices. In both cases, the vorticity increases before the downflow and the flux density, suggesting that the vortex occurs first and convective collapse follows. Further studies are needed to investigate whether this behavior is common to all magnetic structures of the NE. 

Our horizontal velocity map sequence shows many other instances of vortex flows along the boundary of the same supergranular cell. They are significantly shorter than the event studied here (on the order of a few hours), and therefore they are not detected in Figure \ref{fig:fov}(a) as they are diluted by the 24\,h average. Nearly all are associated with NE magnetic patches 
that evolve as described above, that is, they strengthen when caught by 
the vortices and weaken when released from them. For these 
reasons, we are tempted to speculate that the very existence of NE 
magnetic elements, and thus of the network, is due to persistent vortices at supergranular junctions. This scenario is supported by kinematic models of supergranular diffusion \citep{Simon95}, where magnetic fields form isolated clumps as they accumulate at sinks that lie between supergranular cells. 

Photospheric magnetic vortices have been suggested as an efficient mechanism for channeling energy from the lower into the upper solar atmosphere \citep{Wedemeyer12}.  Based on numerical simulations, \citet{Wedemeyer12} found that “magnetic tornado” may carry a significant upward Poynting flux. A solar tornado is generated in the photosphere when a magnetic structure is caught by a vortex flow \citep{Wedemeyer14}. In this layer, magnetic fields are frozen into the plasma and the entire magnetic structure corotates with the vortex. In the chromosphere, the plasma-$\beta$ (i.e., the ratio of gas pressure to magnetic pressure) decreases and the plasma is forced to follow the rotation of magnetic fields. This can produce observational imprints in the chromosphere,
such as the  swirling motions found by \citet{Wedemeyer09} in the core of the Ca\,\textsc{ii} line at 854.2\,nm. If these chromospheric swirls are indeed produced by magnetic tornadoes, then they should appear cospatial to magnetic vortex flows in the photosphere. The confirmation of the  magnetic tornado scenario thus requires the simultaneous detection of photospheric magnetic vortices, similar to the one reported here (Figure \ref{fig:fov}d), and their chromospheric counterparts. Such observations could be carried out, for example, with the CRisp Imaging SpectroPolarimeter \citep{Scharmer08}, where the Ca\,\textsc{ii} line at 854.2\,nm can be measured together with imaging spectropolarimetry in the photospheric Fe\,\textsc{i} lines at 630.2 and 617.3\,nm \citep[see, e.g.,][]{Ortiz14,Rouppe16}. In addition, a method able to estimate the horizontal plasma velocities at every time step is desirable, since vortex flows are not always as persistent as in here, and they can indeed be very ephemeral in IN areas. An end-to-end deep neural network that infers instantaneous horizontal plasma velocities has recently been designed by \citet{Asensio17}. The network is able to detect as small and short-lived vortex flows as those found in simulations, with lifetimes of only a few minutes and diameters of $\sim$\,100 km \citep{Moll11}. This  technique together with the adequate data set may shed light on the nature of magnetic tornadoes.


\begin{acknowledgements}
This paper is based on data acquired in the framework of the \textit{Hinode} Operation Plan 151. We thank the \textit{Hinode} Chief Observers for the efforts they made to accommodate our demanding observations. \textit{Hinode} is a Japanese mission developed and launched by ISAS/JAXA, with NAOJ as a domestic partner and NASA and STFC (UK) as international partners. It is operated by these agencies in co-operation with ESA and NSC (Norway). This work has been partially funded by the Spanish Ministerio de Econom\'{i}a y Competitividad through Projects No. ESP2013-47349-C6-1-R, ESP2014-56169-C6-1-R, and ESP2016-77548-C5-1-R, including a percentage from European FEDER funds. The research leading to these results has received funding from the European Union’s Horizon 2020 programme under grant agreement no. 739500 (PRE-EST project). This research has made use of NASA’s Astrophysics Data System Bibliographic Services.  We acknowledge the community effort devoted to the development of the following open-source packages that were used in this work: numpy
(numpy.org) and matplotlib (matplotlib.org).
\end{acknowledgements}

%
%

\bibliographystyle{aa}
\bibliography{magnetic_vortex.bib}

\end{document}